\documentclass[aps,prd,twocolumn,superscriptaddress]{revtex4}
\usepackage{graphicx}
\usepackage{amsmath}
\usepackage{epsfig}
\usepackage{amsmath,amsfonts,amssymb}
\usepackage{color}

\usepackage[colorlinks]{hyperref}

\begin{document}
\baselineskip=12pt
\def\black{\textcolor{black}}
\def\red{\textcolor{red}}
\def\blue{\textcolor{blue}}
\def\green{\textcolor{black}}
\def\beq{\begin{equation}}
\def\eeq{\end{equation}}
\def\bea{\begin{eqnarray}}
\def\eea{\end{eqnarray}}
\def\orc{\Omega_{r_c}}
\def\om{\Omega_{\text{m}}}
\def\E{{\rm e}}
\def\bearst{\begin{eqnarray*}}
\def\eearst{\end{eqnarray*}}
\def\peleven{\parbox{11cm}}
\def\peffec{\peight{\bearst\eearst}\hfill\peleven}
\def\pspace{\peight{\bearst\eearst}\hfill}
\def\ptwelve{\parbox{12cm}}
\def\peight{\parbox{8mm}}


\title{Transient Weak-Lensing by Cosmological Dark Matter Microhaloes}

\author{Sohrab Rahvar}
\email{srahvar@pitp.ca}
\address{Perimeter Institute for Theoretical Physics, 31 Caroline St. N., Waterloo, ON, N2L 2Y5,Canada}
\address{Department of Physics, Sharif University of
Technology, P.O.Box 11365--9161, Tehran, Iran}

\author{Shant Baghram}
\email{baghram@ipm.ir}
\address{Department of Physics and Astronomy, University of Waterloo, 200 University Avenue West, Waterloo, ON, N2L 3G1, Canada}
\address{Perimeter Institute for Theoretical Physics, 31 Caroline St. N., Waterloo, ON, N2L 2Y5,Canada}
\address{School of Astronomy, Institute for Research in Fundamental Sciences (IPM),P.~O.~Box 19395-5531,Tehran, Iran}

\author{Niayesh Afshordi}
\email{nafshordi@pipt.ca}
\address{Department of Physics and Astronomy, University of Waterloo, 200 University Avenue West, Waterloo, ON, N2L 3G1, Canada}
\address{Perimeter Institute for Theoretical Physics, 31 Caroline St. N., Waterloo, ON, N2L 2Y5,Canada}

 \vskip 1cm

\begin{abstract}
We study the time variation of the apparent flux of cosmological point sources  due to the transient weak lensing by dark matter
microhaloes. Assuming a transverse motion of microhaloes with respect
to our line of sight, we derive the correspondence between the temporal power spectrum of the weak lensing magnification, and the spatial power spectrum of density on small scales. Considering different
approximations for the small scale structure of dark matter, we
predict the apparent magnitude of cosmological point sources to
vary by as much as $10^{-4}-10^{-3}$, due to this effect, within a period of a
few months. This red photometric noise has an almost perfect gaussian statistics, to one part in $\sim 10^4$. We also compare the transient weak lensing power spectrum with the background effects such as the stellar microlensing on cosmological scales. A
quasar lensed by a galaxy or cluster like SDSSJ1004+4112 strong lensing system, with multiple images, is a
suitable system for this study as: (i) using the time-delay method between different images, we can remove the intrinsic variations of the quasar, and (ii) strong lensing enhances signals from the transient weak lensing. We also require the images to form at large
angular separations from the center of the lensing structure, in order to  minimize contamination by the stellar microlensing. With long-term monitoring of quasar strong lensing systems with a 10-meter class telescope, we can examine the existence of dark microhaloes as the building blocks of dark matter structures. Failure to detect this signal may either be caused by a breakdown of cold dark matter (CDM) hierarchy on small scales, or rather interpreted as evidence against CDM paradigm, e.g. in favor of modified gravity models.\\


\end{abstract}

\maketitle

\newpage

\section{INTRODUCTION}\label{intro}
The standard model of cosmology requires existence of cold dark
matter (CDM) particles. The main role of CDM particles is to establish
the gravitational skeleton of large scale
structures \cite{Frenk:2012ph}. An important statistical property of
the dark matter structures is the power-spectrum of the structures
which is predicted by e.g. the inflationary cosmology as one of the possible
paradigms for the early universe \cite{Guth:1980zm}. The standard
model of cosmology predicts nearly scale invariant, adiabatic and
Gaussian distribution of matter at the early stages of the
universe and the evolution of these structures strongly depend on
the properties of the dark matter particles \cite{Dodelson2003}. \\

One of the predictions of CDM paradigm is the hierarchical
structure formation, the existence of CDM
sub-structures (from sub-haloes, down to microhaloes) embedded in the larger haloes
\cite{benmoore}. The larger haloes host baryonic matter in the form
of hot gas and stars at their centers. However, smaller haloes do not have
enough gravitational potential to confine baryonic matter, hence
microhaloes may not host any baryonic component.

An alternative to the dark matter paradigm is the modified gravity models, where the missing gravitational mass in the structures, and the universe, is
compensated by the modification to the law(s) of gravity
\cite{Milgrom,teves,mog,rah13}. Both dark matter and modified gravity
models, at some level, can explain the rotation curves of the galaxies
and formation of the structures \cite{san99,moffat,sobouti}.
However, it is hard to explain the whole range of observations, as the  
cluster of galaxies in 
Modified Newtonian Dynamics (MOND) \cite{bullet} fails, while the Scalar-Tensor Vector Gravity (STVG) 
also so-called MOG can explain the dynamics of clusters in the context of baryonic matter \cite{rah13b}.

\begin{figure}[t]
\includegraphics[width=\linewidth]{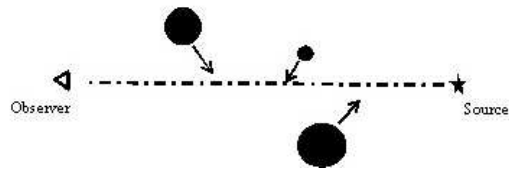}
\caption {The Dark Matter substructures moving relative to our line
of sight towards a cosmological source.} \label{micro}
\end{figure}

The major efforts to test the existence of CDM
particles have been performed in the underground experiments and null
results from these experiments continue to push further constraints on the
parameter space of the dark matter particles (e.g., \cite{Bertone:2004pz, Beltrame:2013bba}). Similarly, the indirect searches for CDM are yet to result in conclusive evidence of its presence. While the main evidence for dark matter comes from its  gravitational effect on large scales, different campaigns have studied its small-scale properties. One of the main efforts has been done by the gravitational microlensing experiments. These experiments
monitor stars in the Large and Small Magellanic clouds and measured
the magnification of stars by lensing. After a decade of observation,
they could rule out dark matter in the form of massive compact
objects in the halo, for nearly the entire range of viable astrophysical masses \cite{eros,macho}.
Another proposal is astrometrical observation of the
background stars by the lensing effect of the microhaloes
\cite{adr}. The substructures in the parent halo can also modify the flux ratio of images in a
strong lensing system \cite{fratio}. The microhaloes may also change the pulsation time of
the pulsars due to the Shapiro and Doppler effects
\cite{Baghram:2011is}.

Here, we propose yet another indirect gravitational method for the possible observation
of the dark matter microhaloes, using the transient gravitational
lensing on cosmological scales. In the dark matter scenario, the
hierarchy of the structure formation predicts the existence of microhalo structures embedded in the larger haloes \cite{benmoore}. These structures have formed at high redshifts, and due to their
small masses, have small virial velocity dispersions of
$\lesssim$ km/s. Due to the thermal pressure of the intergalactic medium, the baryonic matter cannot cool and
condense into the gravitational potential of the microhaloes. The
result is the formation of small non-baryonic structures, made
of only the dark matter. In the case that these microhaloes cross our line of sight towards a cosmological source, such as
a quasar, they could cause a slight change in the source flux due to the weak lensing
effect. Unlike to the traditional weak lensing on the cosmological scales, the
motion of the microhaloes provides a time varying magnification
effect.

Let us make a simple estimate of the effect: Observations over a time period $t$ are sensitive to microhaloes of the size $r \sim v t$, where $v \sim 500$ km/s is the characteristic peculiar velocity on cosmological scales. Assuming the hierarchical structure formation scenario, we put all the mass of the universe in microhaloes of density $\Delta \bar{\rho}_m$, i.e.:
\beq
m = \bar{\rho}_m \times \Delta \times \frac{4}{3}\pi r^3.
\eeq
If a typical microhalo forms at redshift $z \sim 20$ (e.g. \cite{Angulo:2009hf}), its density is enhanced by a factor of $\sim 200$ due to collapse, and then diluted by a factor of $\sim 0.03$ \cite{Afshordi2010} due to subsequent tidal stripping, so we find $\Delta \sim 20^3 \times 200 \times 0.03 \sim 10^5$. The magnification by a single microhalo at distance $D$ is then given by:
\bea
&& \delta A_{\rm single} \sim 4\pi G\frac{ m}{\pi r^2}\times D \sim (2\Omega_m H^2_0) (vt) D  \Delta \nonumber\\ \sim  &&5	 \times 10^{-9} \left(v \over 500~ {\rm km/s}\right)\left(t \over 1~ {\rm yr}\right) \left(\Delta \over 10^5 \right)  \left(D \over 3~ {\rm Gpc}\right),\nonumber\\
\eea
where we assumed $\Omega_m=0.3$ and $H_0 = 70$ km/s Mpc$^{-1}$. This is clearly a tiny number! However, let us now estimate the optical depth, or the number of such microhaloes that cross the line of sight:
\bea
&&\tau = \frac{ \bar{\rho}_m}{m} \times \pi r^2 \times D \sim \frac{3D}{4vt\Delta} \nonumber\\ \sim && 4\times 10^{7} \left(v \over 500~ {\rm km/s}\right)^{-1}\left(t \over 1~ {\rm yr}\right)^{-1} \left(\Delta \over 10^5 \right)^{-1}  \left(D \over 3 ~{\rm Gpc}\right). \nonumber\\
\eea
Therefore, the variance or noise in magnification due to the {\it collective} random contributions of many microhaloes becomes:
\bea
&&\delta A_{\rm tot.} = \sqrt{\tau} \delta A_{\rm single}\sim  \nonumber\\ && 3 \times 10^{-5} \left(v \over 500~ {\rm km/s}\right)^{1/2}\left(t \over 1~ {\rm yr}\right)^{1/2} \left(\Delta \over 10^5 \right)^{1/2}  \left(D \over 3 ~{\rm Gpc}\right)^{3/2}.\nonumber\\
\eea
This simple estimate illustrates the magnitude of the red stochastic photometric noise expected from transient weak lensing by millions of microhaloes that cross the line of sight to cosmological point sources. As we show in the paper, this effect could be magnified by as much as $\sim 30$ for strongly lensed quasars, leading to variabilities of $\delta A/A \sim 10^{-3}$ over months to years timescales. Moreover, the signal is expected to be close to gaussian, roughly to one part in $\sqrt{\tau} \sim 10^4$ (from the central limit theorem), as it is contributed by many uncorrelated (and comparable) microhaloes. Therefore, temporal power spectrum should provide all the statistics necessary to quantify this noise.

One may worry that the cut-off of the CDM hierarchy is ignored in the above estimate, and no variation will be observed below a time-scale associated with this cut-off. In particular, the time-scale to cross the virial radius of a $10^{-6} M_{\odot}$ microhalo that formed at $z\sim 20$  is $\sim 40$ years \footnote{We thank James Taylor for bringing this point to our attention.}. Nevertheless, even the smallest haloes are predicted to have a central cusp in the CDM paradigm, and thus their passage close to the line of sight can lead to rapid time-variations. For example, the cross-section for being within distance $r<r_{vir}$ of  a central cusp is $\propto r^2$, while the surface density for a singular isothermal profile is $\Sigma \propto r^{-1}$. Since $\delta A_{\rm tot.} = \sqrt{\tau} \delta A_{\rm single}$, these two effects exactly cancel each other. Simulated microhalo profiles are shallower in their center than singular isothermal  (e.g. $\Sigma \propto r^{-0.4}$ in \cite{Anderhalden:2013wd}),   which leads to a slight steepening of the transient weak lensing noise below time-scales associated with the scale-radius of microhaloes. We further quantify this effect in Sec. \ref{SDSS}.    

In the rest of the paper, we present a precise formulation for the variability power spectrum in the lensing magnification, by assigning a velocity
field to the perturbation of the metric perpendicular to our line of
sight towards a given quasar. We do this calculation, both using the
geodesic equation and using luminosity distance corrections to the
focusing equation. Finally, we compare results with the present data
from quasars and suggest an observational strategy for long term light curve
measurement of quasars with high precision photometry.

The structure  of paper is as follows: In Section \ref{wl} we use
the perturbation theory in FRW and obtain fluctuations in the
flux of quasar due to weak lensing of microhaloes on the
cosmological scales. In Appendix \ref{angd} we repeat this
calculation with the angular diameter distance formalism. We obtain
the expected temporal power spectrum of the fluctuations of the light
curve of a quasar for different non-linear models of structures in
Section \ref{obs}.  We then compare microhalo weak lensing signals
with the intrinsic variation of quasar light and other backgrounds,
such as stellar microlensing on the cosmological scales. In particular, we propose monitoring of multiply imaged quasar strong lensing systems as a way to distinguish  the transient weak lensing effect from intrinsic variations. The conclusions are given in Section \ref{conc}.


\section{Transient Weak Lensing in the perturbed Universe: Geodesics Method}
\label{wl} In this section, we formulate the time variation of a
source flux located at a cosmological distance, due to the lensing
effect of DM substructures.  We let the perturbations in the FRW
metric move relative to the line of sight due to the peculiar
velocity of structures and dispersion velocity of sub-structures
embedded in a larger structure. Hence the transverse velocity
produce a transient weak lensing effect and observationally the
light curve of source can change with time.

We start with calculation of the geodesics equation of the light
ray from the source to the observer. The perturbed FRW metric in the Newtonian gauge
for the matter dominant era is given as below \cite{bardeen}:
\begin{equation}
ds^2=-[1-2\Phi(\vec{x},t)]dt^2+a^2\delta_{ij}[1+2\Phi(\vec{x},t)]dx^i
dx^j, \label{pert-met}
\end{equation}
where we set $c=1$. The transverse spatial component of the light
ray as well as the longitudinal direction follows the geodesics
equation:
\begin{equation}
\frac{d^2x^{i}}{d\lambda ^2}+\Gamma
^{i}_{\mu\nu}\frac{dx^{\mu}}{d\lambda}\frac{dx^{\nu}}{d\lambda}=0,
\end{equation}
where $i$ is the index of the spatial coordinate and $\lambda$, the affine parameter is the comoving time
measured by the observers along the light ray. This affine parameter
in the homogenous FRW universe is related to the coordinate time $t$
and comoving distance $\chi$ through the scale factor:
\begin{equation}\label{affine}
d\lambda=adt=a^2d\chi.
\end{equation}
We assume a small transverse perturbation in the trajectory of the
light rays compare to the longitudinal observer-lens and
observer-source distances. Using the Christoffel symbols from metric
in Eq.(\ref{pert-met}) and ignoring the higher orders of the
perturbations, the geodesic equation simplifies to
\begin{equation}\label{geo1}
\frac{d^2x^{i}(\chi)}{d\chi^2}-2\Phi_{,i}=0,
\end{equation}
where we replace the affine parameter with the comoving distance
form Eq. (\ref{affine}). Now we change the transverse comoving
coordinate with the angular position of the light ray to get the
angular position of the source as function of angular position of images. Integrating
Eq.(\ref{geo1}), we obtain
\begin{equation}\label{lens1}
\beta ^{i}=\theta ^{i}+
\frac{2}{\chi_s}\int_{0}^{\chi_{s}}d\chi^{\prime\prime}\int_{0}^{\chi^{\prime\prime}}\Phi_{,i}(\chi
^{\prime})d\chi ^{\prime},
\end{equation}
where $\beta^{i}$ is the angular position of the source,
$\theta^{i}$ is the observed angular position of the image and
$\chi_{s}$ is the comoving distance of the observer to the source.
We simplify the double integral as
\begin{equation}\label{lens1}
\beta ^{i}=\theta ^{i}+ {2}\int_{0}^{\chi_{s}}\Phi_{,i}(\chi
^{\prime})\left(1-\frac{\chi^\prime}{\chi_s}\right)d\chi ^{\prime}.
\end{equation}
The maping matrix from the image space to the source space is given
by
\begin{equation}\label{transform}
A_{ij}=\frac{\partial\beta ^{i}}{\partial\theta ^{j}}=\delta
_{ij}+2\int\Phi_{,ij}(\chi
^{\prime})\left(1-\frac{\chi^{\prime}}{\chi_s}\right)\chi^{\prime}d\chi^{\prime}.
\end{equation}
The Jacobian of transformation matrix in Eq.(\ref{transform}) within
the framework of geometric optics provides the magnification by $A =
1/det(A_{ij})$. For the low magnifications, ignoring higher order terms, $A$
is given by
\begin{equation}\label{mag1}
A\simeq1-2\int\nabla_{2D}^2\Phi(\chi^{\prime})\left(1-\frac{\chi^{\prime}}{\chi_s}\right)\chi^{\prime}d\chi
^{\prime}.
\end{equation}
We note that $\nabla_{2D} ^2$ is defined in two dimension lens
plane, perpendicular to the line of sight. In the Fourier space
$\nabla ^2_{2D}\Phi(\chi^{\prime},t)$ can be written as
\begin{equation}\label{fourier}
\nabla ^2_{2D}\Phi(\chi^{\prime},t)=-\int
\frac{d^3k}{(2\pi)^3}k_{\bot}^2\Phi(k)e^{-ik_{\bot}\chi_{\bot}(t)}e^{-ik_{||}\chi_{||}},
\end{equation}
where $k_{\bot}$ and $k_{||}$ are  transverse and parallel wavenumbers.
In writing Eq. (\ref{fourier}), we use the so-called {\it moving sheet} approximation \cite{Baghram:2011is}, where we assume the time dependence of the lensing potential is entirely due  to the coherent transverse motion of a  lensing sheet. This approximation is justified as motion is dominated by cosmic velocities that are coherent on large scales.
   Therefore, in Eq. (\ref{fourier}) , only the transverse coordinate is a function of time, $\chi_\bot =
\chi_\bot(t)$. Substituting Equation (\ref{fourier}) in
(\ref{mag1}), the time variation of the magnification $A$ obtained
as:
\begin{eqnarray}
\delta A(t)&=&2\int_{0}^{\chi_{s}}\left(1-\frac{\chi
^{\prime}}{\chi_s}\right)\chi^{\prime}d\chi^{\prime}\nonumber\\
&\times& \int \frac{d^3k}{(2\pi)^3}
k_{\bot}^2\Phi(k)e^{-ik_{\bot}\chi_{\bot}(t)-ik_{||}\chi^{\prime}}
.\label{permag}
\end{eqnarray}
In Appendix \ref{angd}, we repeat this calculation and derive
magnification variation based on the angular diameter distance
method. Now using this result, we investigate the statistics of
magnitude variation as a function of duration of observation in the
next section.

\section{Characteristics of quasars light curve: Transient weak lensing}

\label{obs} Imagine that we are measuring the light curve of a point
like source at a cosmological distance. For an ideal quasar with
stable light curve, we can define the correlation function in the
magnification of the source due to the lensing by the microhaloes as
follows:
\begin{widetext}
\begin{eqnarray}\nonumber
\langle\delta A(t_1)\delta A(t_2)\rangle &=& 4\int_{0}^{\chi_{s}}
\left(1-\frac{\chi^{\prime\prime }}{\chi_{s}}\right)\chi^{\prime\prime}
d\chi^{\prime\prime} \int_{0}^{\chi_{s}}\left(1-\frac{\chi^{\prime
}}{\chi}_{s}\right)\chi^{\prime} d\chi^{\prime}\int
\frac{d^3k}{(2\pi)^3}\frac{d^3k'}{(2\pi)^3}\int \frac{vdv}{\sigma
^2}\left({k_{\bot}k'_{\bot}}\right)^2\langle\Phi(k,z)\Phi(k',z)\rangle\\
\label{cor1} &\times& \exp\left[-ik_{||}\chi^{\prime}-ik^{\prime}_{||}\chi^{\prime\prime} -v^2/2\sigma^2-ivt_1
k_{\bot}/a-ivt_2k^{'}_{\bot}/a\right],
\end{eqnarray}
where $t_1$, $t_2$ correspond to two subsequent observations with the
interval of $\tau = t_1 - t_2$ and $v$ is the transverse velocity of
substructures, which we have assumed to have a Maxwellian distribution and 1d dispersion velocity
of $\sigma$. Here the transverse velocity is normalized to the
cosmological scale factor to represent the velocity in the comoving
space. We note that the peculiar velocity of the structures is a
redshift and scale dependant parameter. Using the definition of the
potential power spectrum:
\begin{equation}
\langle\Phi(\vec{k},z)\Phi(\vec{k'},z)\rangle=(2\pi)^3\delta^{3}(\vec{k}+\vec{k'})P_{\Phi}(\vec{k},z),
\end{equation}
and substituting in Eq.(\ref{cor1}), we can integrate over
$k^{\prime}$ for an ensemble of structures as follows:
\begin{equation}
\langle\delta A(t_1)\delta A(t_2)\rangle =
4\int_{0}^{\chi_{s}}(1-\frac{\chi^{\prime\prime
}}{\chi_{s}})\chi^{\prime\prime}
d\chi^{\prime\prime}\int_{0}^{\chi_{s}} (1-\frac{\chi^{\prime
}}{\chi_{s}})\chi^{\prime}d\chi^{\prime}\int \frac{vdv}{\sigma
^2}\int\frac{d^2k_{\bot}}{(2\pi)^2}\frac{dk_{||}}{2\pi}k_{\bot}^4P_{\Phi}(\vec{k},z)
e^{-ik_{||}(\chi^{\prime}-\chi^{\prime\prime})}e^{-i(t_1-t_2)vk_{\bot}/a}e^{-v^2/2\sigma^2}.
\end{equation}
Ignoring the longitudinal contribution in $P(k,z)$ (i.e. $k_{\bot} \gg k_{||}$), and integrating
over $k_{||}$ and $\chi^{\prime\prime}$, the correlation function
simplifies to:
\begin{equation}\label{cor2}
\langle\delta A(t_1)\delta A(t_2)\rangle =4\int
_{0}^{\chi_{s}}(1-\frac{\chi^{\prime}}{\chi_{s}})^2 {\chi
^{\prime}}^2 d\chi^{\prime} \int \frac{vdv}{\sigma ^2}\int
\frac{dk_{\bot}}{2\pi} P_{\Phi}(k,z) k^5_{\bot}
e^{-ivk_{\bot}(t_1-t_2)/a}e^{-v^2/2\sigma^2}.
\end{equation}
In order to relate this correlation function with the observation,
we replace the power spectrum in the potential with the density
contrast, using the Poisson equation:
\begin{equation}
k^2\Phi(k,z)=4\pi G\rho_m(z)\delta(k,z)(1+z)^{-2},
\end{equation}
where $\rho_m$ is the density of universe at redshift $z$. Using FRW
equation, we can we write the Poisson equation as
\begin{equation}
\Phi(k,z)=\frac{3H_{0}^2\Omega_{m}^{0}(1+z)}{2k^2}\delta(k,z).
\end{equation}
Consequently, the power spectrum of the potential in terms of
dimensionless power-spectrum $\Delta^2(k,z)=|\delta_k|^2k^3/2\pi^2$
is given by
\begin{equation}\label{phipot}
P_{\Phi}(k,z)=\frac{9\pi^2}{2k^7}H^4_{0}{\Omega^{(0)}_{m}}^2(1+z)^2\Delta^2(k,z).
\end{equation}
Substituting Eq.(\ref{phipot}) in Eq.(\ref{cor2}), correlation function of the light curve is
given by
\begin{equation}\label{cor3}
 \langle\delta A(t_1)\delta
A(t_2)\rangle =18\pi ^2H^4_{0}{\Omega_{m} ^{(0)}}^2\int
_{0}^{\chi_{s}}(1-\frac{\chi ^{\prime}}{\chi_{s}})^2{\chi
^{\prime}}^2 d\chi ^{\prime} \int \frac{vdv}{\sigma ^2}\int
\frac{dk_{\bot}}{2\pi} \frac{\Delta ^{2}(k,z)}{
k^2_{\bot}}(1+z(\chi'))^2e^{-ik_{\bot}v(t_1-t_2)/a}e^{-v^2/2\sigma^2}.
\end{equation}
For structure with the size of $k^{-1}a$, a given transverse
velocity of $v$ implies the frequency of $\omega=k_{\bot}{v}/a$ in the light curve.
Replacing $k_{\bot}$ with $\omega$, we use the definition of the
temporal power spectrum as follows
\begin{equation}
\langle\delta A(t_1)\delta A(t_2)\rangle=\frac{1}{2\pi}\int
P(\omega)e^{-i\omega\tau}d\omega, \label{tpow}
\end{equation}
where we replace $\tau=t_1-t_2$. Then the dimensionless power spectrum of the
light curve is given by
\begin{equation}\label{pomega}
\omega P(\omega)=18\pi^2H^4_{0}{\Omega^{(0)} _{m}}^2
\int_{0}^{\chi_{s}} \left(1-\frac{\chi ^{\prime}}{\chi_{s}}\right)^2{\chi
^{\prime}}^2 d\chi^{\prime} \int_{0}^{\infty}{dv}~
e^{-v^2/2\sigma^2}\left[\frac{v}{\sigma(\frac\omega v,z_{\chi'})
}\right]^2 \frac{\Delta^2(\frac{\omega a}
{v},z_{\chi'})}{\omega}(1+z_{\chi^{\prime}})^{{3}},
\end{equation}
\end{widetext}
Here the power spectrum of magnification depends on distribution of
DM which manifests itself by the matter power spectrum
$\Delta^2(k)$. In order to calculate power-spectrum $P(\omega)$,
corresponding to the light curve, we need a model for the evolution of
velocity dispersion of DM substructures and the distribution of the
matter in the small scales.

The velocity of CDM microhaloes results from the dispersion velocity
in the galactic haloes, which in the host halo of typical galaxies
$\sigma\sim 200$ km/s, and from the bulk flow velocity in the
intergalactic medium, that is $\sigma\sim 500$ km/s
\cite{Abate2011,Sheth2001}. Now, in order to estimate the sensitivity of our
observations to the scale of the relevant dark matter structures, we assume performing the light curve observations of a quasar from $1$ day to $10$ years.
Assuming the corresponding velocities for the microhaloes, the time
variation of quasar will probe nonlinear regime on the length scales
of $k\sim10 ^{8} -10^{12}$ Mpc$^{-1}$. These scales are
significantly smaller than the reach of numerical simulations, or
standard semi-analytical models (e.g. halo model) \cite{Ma:2000ik}.

In the next subsection, we review the stable clustering hypothesis
as a theoretical model for the small scale nonlinear clustering of
CDM (and its microhaloes). We then use this model to predict the
amplitude of fluctuations in the quasar light curves due to
transient weak lensing.

\subsection{Nonlinear matter power Spectrum: Stable Clustering in Phase Space}
In this section we introduce the semi-analytical formalism of
stable clustering in the phase space, along with other non-linear models of small scale clustering. The
reason we require a semi-analytic model such as stable clustering is that numerical simulations
can only explore the distribution of matter up to the scales of
$k\sim10^{3}$ Mpc$^{-1}$ \cite{Bolyan2009}. Moreover, currently there is
no reliable observational constrains on the CDM power spectrum on sub-Mpc small
scales.

\begin{figure}[t]
\includegraphics[width=\linewidth]{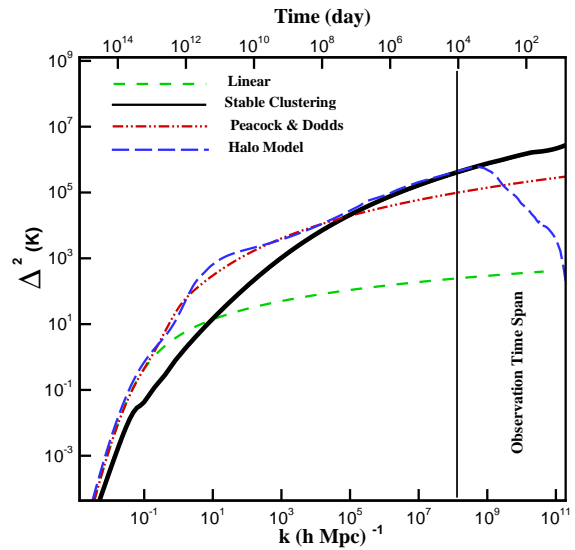}
\caption {Dimensionless power spectrum of density fluctuations
$\Delta^2(k) = k^3 P_{NL}(k)/ (2\pi^2)$ as a function of wavenumber
$k$ for the linear regime (dashed line), Peacock and Dodds fitting
formula (dash-dot-dot line), halo model (long-dashed line) and
stable clustering hypothesis  (solid line) with $\mu=0.1$} \label{figPS}
\end{figure}

The stable clustering model was first introduced by Davis \& Peebles as an analytical
method for calculating the correlation function of the structures in
the deep non-linear regime \cite{dp}. In this model it is assumed
that the number of neighboring particles remains fixed. In other words, the pairwise
velocity vanishes on small scales for non-linear structures. The idea was later extended to the phase space, where the number of particles in the
vicinity of each point (i.e. fixed position and velocity) was assumed to not change with time \cite{Afshordi2010,Zavala:2013bha}.
In the stable clustering hypothesis in phase space, for the small
scales, the correlation function of densities is related to
phase-space density as \cite{Baghram:2011is}:
\begin{eqnarray} {\label{Eqcorr}}
\langle\rho(\vec{r}_{{\i}})\rho(\vec{r}_{{\i}{\i}})\rangle&=&\int
d^3\vec{v_{{\i}}}d^3\vec{v_{{\i}{\i}}}\langle f(\vec{r}_{{\i}},\vec{v}_{{\i}})f(\vec{r}_{{\i}{\i}},\vec{v}_{{\i}{\i}})\rangle\\
\nonumber &\simeq&\int d^3\bar{\vec{v}}d^3\Delta \vec{v} \mu \langle
f(\vec{\bar{r}},\vec{\bar{v}})\rangle\xi_{s}(\Delta r,\Delta v)\\
\nonumber &=&\mu\bar{\rho}_{avg}\int
d^3\Delta\vec{{v}}\cdot\xi_{s}(\Delta {r},\Delta v).
\end{eqnarray}
where $f$ is the phase space density, $\bar{\rho}_{avg}$ is the
average density of matter and $\mu$ is the fraction of dark matter
substructure that survived tidal stripping. $\xi_s$ is the phase
space density of DM particles in the small volume of phase space
($\Delta v$ and $\Delta r$) obtained from spherical
collapse results \cite{Gunn:1972sv} and is related to the linear matter
power spectrum through the variance of matter density (see \cite{Afshordi2010} for a
detailed calculations). For our
study, we also need to know  the matter power spectrum  at different
redshifts. Within the stable clustering hypothesis, the phase-space density $\xi_s$
is assumed to be constant, and thus the evolution is simply given by the
evolution of the background density. Hence the correlation function
for the the small size non-linear overdensities can be written as:
\begin{equation}\label{SC}
\langle\delta(\vec{r}_{{\i}})\delta(\vec{r}_{{\i}{\i}})\rangle\sim\frac{
\mu}{\bar{\rho}_{avg}(1+z)^3}\int d^3\Delta\vec{{v}}\cdot\xi_{s}(\Delta
{r},\Delta v).
\end{equation}
In Fig.(\ref{figPS}), we plot the dimensionless matter power-spectrum
assuming stable clustering hypothesis in the phase space and
$\bar{\rho}_{avg}=\rho _{\rm crit.}$ \footnote{In the case of microhaloes in a
background parent halo, $\bar{\rho}_{avg}$ should be the density of
the halo that hosts the microhaloes}, where $\mu=0.1$ and the minimum mass of
subhaloes is set to $M_{DM}=10^{-12}M_{\odot}$. In Fig.(\ref{figPS}),
we also compare the fitting formula of Peacock and Dodds
\cite{Peacock1996} (dash-dot-dot line) and the Halo model
\cite{Smith2003} (long dash line) for the non-linear power
spectrum with the stable clustering. Now from the matter power
spectrum and the assumption for the velocity dispersion of DM haloes
we can compute the temporal power spectrum of quasar light curve.


%

\begin{figure}[t]
\includegraphics[width=\linewidth]{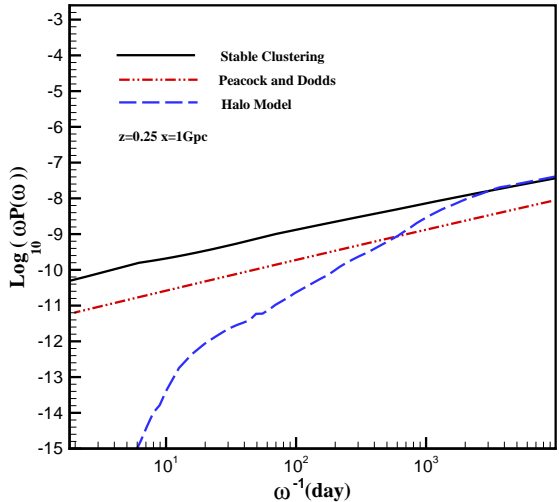}
\caption {Dimensionless power spectrum of magnification from
Eq.(\ref{pomega}) is plotted versus $\omega^{-1}$. The spectrum is
plotted for stable clustering hypothesis (solid line) with $\mu=0.1$, Peacock and
Dodds fitting function (dashed-doted line) and for halo model (long
dashed line) with  the cut-off  related
to the size of smallest halo mass of $M_{min} = 10^{-6}M_{\odot}$. Here the source is located at $\chi = 1$ Gpc
($z\sim0.25$).} \label{figPL}
\end{figure}

We note that in Eq. (\ref{pomega}), the velocity of structure is a
combination of substructure velocity embedded in the halo and
velocity of parent structure.
Hence, the overall dispersion velocity is given by:
\begin{equation}\label{dispersion}
\sigma=(\sigma ^2_{vir}+\sigma ^2_{halo})^{1/2},
\end{equation}
where $\sigma_{halo}$ is the dispersion velocity of parent halo
which is determined through linear perturbation theory and
$\sigma_{vir}$ is the dispersion velocity of sub-haloes in parent
halo. Assuming the spherical collapse model, $\sigma_{vir}$ can be
related to the mass of the parent halo as \cite{bryan}:
\begin{equation}
\sigma_{vir}=476g_{\sigma}(\Delta_{nl}E^2)^{1/6}\left(\frac{m}{10^{15}M_{\odot}/h}\right)^{1/3}km/s,
\end{equation}
here $g_{\sigma}=0.9$, $\Delta_{nl}=18\pi ^2+60x-32x^2$ with
$x=\Omega_{m}-1$, $\Omega_{m}(z)=\Omega^{0}_{m}(1+z)^3$,
$E(z)=\Omega^{0}_{m}(1+z)^3+\Omega_{\Lambda}$. In the linear theory
of structure formation, $\sigma$ in Eq.(\ref{dispersion}) is related
to linear matter power spectrum via
\begin{equation}
\sigma_{halo}(m,z)=Hf\sigma_{-1}\sqrt{1-\sigma^4_{0}/\sigma^2_{1}\sigma^2_{-1}},
\end{equation}
where $f\equiv {d\ln\delta}/{d\ln a}$ is the growth function and the
linear moments of the dispersion velocity are defined by
\begin{equation}
\sigma^2_{j}(m)=\frac{1}{2\pi ^2}\int dk k^{2+2j}P_m(k)W^2[kR(m)],
\end{equation}
where $W(x)=(3/x^3)\left[\sin(x)-x\cos(x)\right]$ is the Fourier
transform of the spherical top-hat filter, and $P_{m}(k)$ is the matter power spectrum. The dispersion velocity
depends both on cosmology, the shape of power spectrum and the
environment of the sub-halo. Using the dispersion velocity of
sub-haloes and theoretical power spectrum in Fig.(\ref{figPS}), we
plot the dimensionless power spectrum of the magnification in equation
(\ref{pomega}) versus the frequency for various nonlinear models in Fig.(\ref{figPL}).

\begin{figure}[t]
\includegraphics[width=\linewidth]{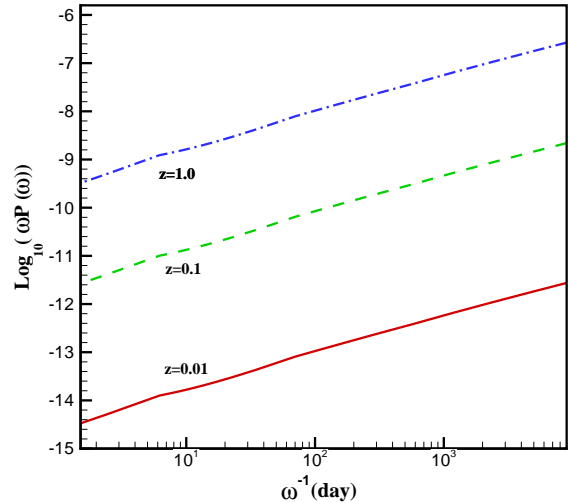}
\caption {Dimensionless power spectrum of magnification from
Eq.(\ref{pomega}) is plotted versus the frequency $\omega^{-1}$ for
sources at different redshifts ( solid line for z=0.01, dashed line
for z=0.1 and dashed dot line for z=1) for stable clustering model with $\mu=0.1$.} \label{fig1}
\end{figure}

For the stable clustering model, we also plot dimensionless power
spectrum $\omega P(\omega)$ for sources in various redshifts in
Fig.(\ref{fig1}). Increasing the redshift of source enhances the
power spectrum of the light curve. Integrating over the power
spectrum, we obtain the variance in the light curve within the
window of $\tilde\omega\in[\tilde\omega_0,\infty]$ which corresponds
to $T\in[0,T_0]$,
\begin{equation}\label{sigma}
\sigma ^2(\omega_{0}; z)=\frac{1}{2\pi}\int_{\omega_{0}}^{\infty}
P(\omega)d\omega.
\end{equation}

If $T_{obs}$ is the duration of observations, then the longest observable mode is constrained
by $k^{-1}<v T_{obs}$ or $\omega > T_{obs}^{-1}$.

\subsection{Observational Target and Possible Backgrounds}
Long term observations of quasars by the MACHO group \footnote{http://wwwmacho.anu.edu.au} during the
monitoring of Magellanic Clouds for detecting Massive Astrophysical
Compact Halo Objects (MACHOs), provided a unique set of quasar light
curves for duration of almost one decade \cite{gela}. They could
report $47$ quasars in the redshift range of $0.2<z<2.8$ with
sampling rate of $2$ to $10$ days. The COSMOGRAIL project 
is also an optical monitoring campaign that aims to measure time delays for a large number of 
gravitationally lensed quasars to accuracies of a few percent using a network of 1- and 2-m class 
telescopes \cite{cosmograil}. The analysis of the light curves
shows that the power spectrum of quasars follow a power law function
as $P(\omega)\propto 1/\omega^2$ \cite{kelly}. This time variation
of the quasar flux can be simulated by the autoregressive process.
In this process each data point on the light curve relates to the
next one by the equation of $F_i = \alpha_{AR}F_{i-1} + \epsilon_i$,
where $\epsilon_i$ is a normally distributed random variable with
zero mean and variance of $\sigma_{AB}$ and
$|\alpha_{AR}| < 1$ in order to ensure stationarity. The physics
behind the variability of the quasar light curve is not clear,
however it can be due to the disk instability \cite{kawa},
microlensing by the intervening stars \cite{wams} or due to the
Poisson processes \cite{fer}. Recently the effect of dark matter
haloes also has been studied in producing the caustic lines in the
lens plane, making small magnification, short duration variations in
the light curve \cite{chen}.

The intrinsic variations of light can be removed if we use the
time delay in the light curve of a quasar, lensed by an intervening
galaxy or cluster of galaxies. The double or multiple images from the strong lensing enable us to find the time delay between the light curves of different images, and eventually intrinsic variations can be removed by time
shifting \cite{timedelay}. While in principle with the time delay
method one can remove the intrinsic variations, there are other
intervening background signals from the microlensing events either by (i) stars intervening light bundles in the lensing galaxy \cite{wam} or (ii) stars belong to the galaxies distributed in the cosmological scales along our line of sight.

The strength of the microlensing signals depend on the column
density of stars along our line of sight. We assume $M_*$ as the
mean mass of stars from the stellar mass function. The Einstein
radius, which characterizes the size of a lens by star is
given by
\begin{equation}
R_E = \sqrt{\frac{4GM_*}{c^2}\frac{\chi(z_l)\chi(z_l,z_s)}{\chi(z_s)}},
\label{einsteinradius}
\end{equation}
where $\chi(z_l)$, $\chi(z_l,z_s)$ and $\chi(z_s)$ are the distances
of observer--lens, lens--source and observer--source in the comoving
frame. High density of stars on the lens plane can produce a network
of caustic lines and crossing quasar with caustic lines produces
strong magnifications in the quasar's light curve. However, the finite-size
of quasar prevents singularities in the light curve and the caustic
crossing of source just produce strong magnifications.

The transition from single occasional lensing to lensing by field of
stars can be quantified by comparing the two dimensional number
density of stars and the Einstein ring of individual stars. Assuming
$n_{2D}$ as the column number density of stars, in order to have
single lensing regime, the relative distance between the lenses must be
larger than the Einstein radius of lenses, $n_{2D}^{-1/2} \gg R_E$.
Substituting the definition of Einstein radius from Equation
(\ref{einsteinradius}), and using the definition of column density
of stars, the condition of occasional single lensing regime implies
$\kappa \ll 1$ where $\kappa = \Sigma(r)/\Sigma_{cr}$ and the
critical column density is defined by
\begin{equation}
\Sigma_{cr} = \frac{c^2}{4\pi
G}\frac{\chi(z_s)}{\chi(z_l)\chi(z_l,z_s)}.
\end{equation}
In what follows we estimate the density of stars in the galactic
halo and provide the condition for occasional
microlensing events. For lens and source located at the cosmological
scales (i.e. $z \simeq 1$), we obtain a critical column density in
the order of $\Sigma_{cr} \simeq 10^3 M_\odot~pc^{-2}$. Now we
assume a galactic halo around the lensing galaxy with isothermal
density of
\begin{equation}
\rho(r) = \frac{M}{4\pi R} \frac{1}{r^2},
\end{equation}
where $M$ is the mass of halo and $R$ is the characteristic size of halo. Assuming a dark halo with a fraction of its mass made of MACHOs, we can relate the density of halo to the density
of MACHOs as $\rho_\star(r) = f \rho(r)$ where from the microlensing
experiments in the direction of Large and Small Magellanic Clouds,
the upper bound of $f$ is $f<0.2$ \cite{lass}. For the case of strong
lensing with a Milky Way size galaxy, $M = 5\times 10^{11} M_\odot$
and $R = 50~$ kpc,  we can calculate the convergence parameter
corresponding to the stars as
\begin{equation}
\kappa_\star = \frac{6\times 10^{-3}}{h} \cos^{-1}(h),
\end{equation}
where $h$ is the position of images of quasars normalized to the
size of halo (i.e. $h = z/R$). For the case of $h< 0.01$, $\kappa_\star\simeq 1$ and we can
detect the caustic crossing features from the network of caustic lines
in the halo structure.
Moreover, for the small impact parameters relative to the center of
galaxy,  the luminous parts of galaxy as disk and bulge also will contribute
in the microlensing. For the outer
regions of halo, $h>0.95$ the convergence parameter is $\kappa_\star < 10^{-3}$ which is the favorable quasar lensing systems that form images around the Einstein ring. For a source, lens and
observer aligned on a straight line, a Milky Way type galaxy as a lens on
cosmological scales, equation (\ref{einsteinradius}) implies that Einstein ring forms at $\simeq 10$ kpc. More massive galaxies or cluster of galaxies can make larger Einstein rings.

Recent analysis of quasar images in the SDSS catalog have found
a sample of 26 lensed quasars brighter than $I = 19.1$ and in the
redshift range of $0.6 < z < 2.2$. These quasars are selected from
$50,826$ spectroscopically confirmed quasars in the SDSS Data
Release 7 (DR7). For this sample of lensed quasars, the image
separation ranges $1^{''} < \theta < 20^{''}$ where the I-band
magnitude differences in two images is smaller than $1.25$ mag
\cite{Naohisa}. This angular separation for the images in the
cosmological scales corresponds to the spatial separation of images
in the rang of $5<L<100$ kpc.

\begin{figure}[t]
\includegraphics[width=\linewidth]{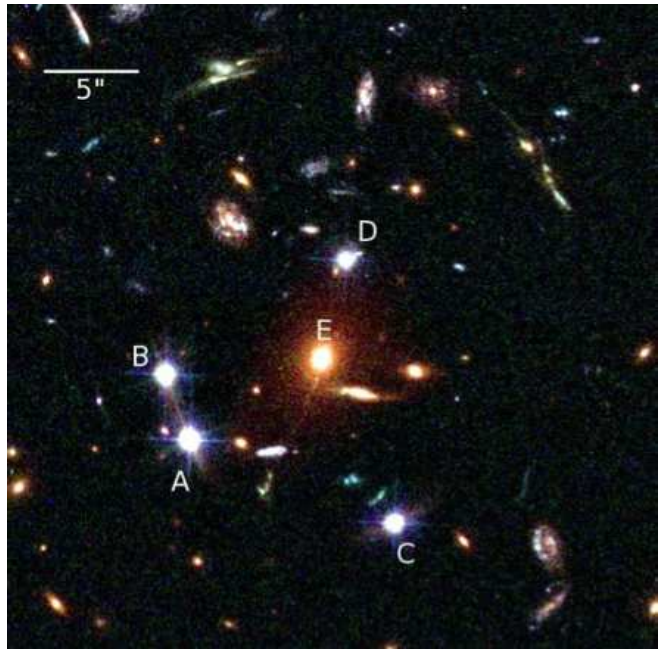}
\caption {SDSSJ1004+4112 strong lensing system as one of the exceptional lensing systems with images are split as large as 14.62 arcsec  separation\cite{inada}. It seems that the lensing system is a cluster of galaxy with concentrated dark matter.  The time delay between the images is of order a year. The apparent magnitude of images in red band are:   $m_i(A) = 18.46\pm 0.02$,  $m_i(B) = 18.86\pm 0.06$,  $m_i(C) = 20.36\pm 0.03$ and  $m_i(D) = 20.05\pm0.04$ \cite{oguri04}.  Photo adapted from \cite{oguri} and by ESA, NASA, K. Sharon (Tel Aviv University) and E. Ofek (Caltech) .} \label{SDSSlensing}
\end{figure}

In some of quasar lensing systems, the lens is a cluster of galaxy with an
extended halo. In this case, the situation is much better than
lensing by a galaxy,  as the haloes of clusters have much less MACHO densities, compared
to the galactic haloes. However, the disadvantage of these systems is that the
time delays between the distant images are longer, and a long-term
survey of quasar's light curve for time-delay analysis is needed. The quasar SDSSJ1004+4112
belongs to this sample \cite{inada}. In this system, we have five images from
the background quasar where some the images are close to the
galaxies and exhibit microlensing features. Figure (\ref{SDSSlensing}) demonstrate the lensing cluster and images from the background quasar. The apparent magnitude of images in red band for these images are:  $m_i(A) = 18.46\pm 0.02$,  $m_i(B) = 18.86\pm 0.06$,  $m_i(C) = 20.36\pm 0.03$ and  $m_i(D) = 20.05\pm0.04$ \cite{oguri04}.  The time ordering between the images are as C-B-A-D-E with the time delays of
$\Delta\tau_{BA} = 40.6\pm 1.8$ days, $\Delta\tau_{CA} = 821.6\pm2.1$ days, $\Delta\tau_{CB} = 681\pm 15$ days and $\Delta\tau_{AD}>1250$ days \cite{timedelay2}.

From the best model to the lensing system at the position of component A, the best values for the convergence and shear are $\kappa = 0.392$ and $\gamma = 0.642$  \cite{Gordon}. Similar to our arguments in the case lensing by a galaxy, images form at larger distances from the center of cluster have least contaminated by the stellar microlensing events \cite{Richards}.

Since most of the trajectory of light traveling from the
source to the observer is in the intergalactic medium, there might
also be microlensing events of cosmological origin as the background events. 
%
%
%
%
 In order to estimate the probability of observation of the
cosmological microlensing events, and their rate, we calculate the
optical depth, or the probability that a background source is located inside
the Einstein ring of a foreground (point) lens \cite{pac}:
\begin{equation}
\tau =  \frac{4\pi G u_0^2}{c^2}\int \bar{ \rho}_\star D_s^2 x(1-x) dx,
\label{optdepth}
\end{equation}
where
$\bar\rho_\star$ is the cosmological mean density of stars
in the universe, 
$u_0$ is the minimum impact parameter (size of lens) normalized to Einstein radius, and $x = D_l/D_s$ is the
ratio of the lens to the source distances. Using the Friedmann
equation we can simplify expression (\ref{optdepth}) as
\begin{equation}
\tau \sim u_0^2 \Omega_\star (H_0D_s)^2,
\label{tau}
\end{equation}
where $\Omega_\star$ corresponds to the fraction of current cosmic density in stars,  and $H_0$ is the Hubble constant. For a quasar at
cosmological distance $D_s \sim H_0^{-1}$. Using 
$\Omega_\star \sim 10^{-2}$,  the optical depth for $u_0 = 1$ (i.e. ${\cal O}(1)$ magnification) 
assuming a uniform distribution of stars on cosmological scales, yields the optical depth of
$\tau\sim 10^{-2}$. A details analysis in \cite{zac07}, assuming that MACHOs follow dark matter content in the background and over--dense parts of university,  confirms our estimated optical depth. This is also consistent with the studies of intracluster light, e.g. in \cite{Budzynski:2013obt}, which find $< 1$\% of the cluster mass to be in the intracluster stars. 

A realistic estimation for a lensing of a quasar by a galaxy can also be obtained by using the distribution function of the galaxies based on their size and number density, throughout the universe. Using the Press-Schechter distribution function \cite{ps}, the optical depth of lensing of objects at the redshift of $z_s\sim 2$ is about $\tau \sim 10^{-2}$ \cite{loeb} which is consistent with what we obtained from Equation (\ref{tau}).
%

While it is reassuring that microlensing events that lead to $>{\cal O}(1)$ magnification are not common, it is clear that magnification events of $\sim 10^{-4}$ have higher optical depth.  The low magnification in the microlensing events is given by $\propto (\theta_E/\theta)^4$ in the field, and $\propto (\theta_E/\theta)^2$ in strongly lensed systems, as function of the angular separation $\theta$, between the quasar and  the star. Therefore, events with $\theta \sim 10 \times \theta_E$ would have optical depth of unity, implying that magnifications of $10^{-4}$ ($10^{-2}$) are common in the field (strongly lensed systems). However, the characteristic time-scale for microlensing events of this size is given by:
\bea
&&t(\theta) \sim \frac{\theta}{\dot{\theta}} \sim \left(\theta\over \theta_E\right) \left(R_E\over v\right)  = \nonumber\\ &&(230 ~{\rm yr}) \left(\theta/\theta_E \over 10\right) \left(M_* \over M_{\odot} \right)^{1/2}   \left(\chi(z_s) \over 3~{\rm Gpc}\right)^{1/2} \left[v({\rm km/s}) \over 500 \right]^{-1},   \nonumber\\ \label{t_theta}
\eea  	
assuming that the typical lens is half-way to the source. Therefore, we see that the characteristic time-scales for typical magnification events (i.e. with $\tau \gtrsim 1$) is quite long, and
thus they are unlikely to contaminate the transient weak lensing due to microhaloes. In Appendix \ref{microlensing}, we derive the power spectrum of weak stellar microlensing light curves, and show that they are suppressed by $\exp[-2 t(\theta)/T_{obs}]$, which is shown in Fig. (\ref{figsigma}) for the nominal parameters in Eq. (\ref{t_theta}).

In denser fields, it is more likely to find a star close to the line of sight, so that $\exp[-2 t(\theta)/T_{obs}]$ is not too small. In such cases, the main microlensing signal will be dominated by the handful of stars within the radius $\dot{\theta}T_{obs}$. Given the small number of lenses, the statistics of the signal will be very non-gaussian, with significant phase correlation among different Fourier amplitudes, as well as a sharp exponential drop at high frequencies (see Fig. \ref{figsigma}). This will be in contrast with the transient weak lensing due to microhaloes, which (as we argued in Sec. \ref{intro}) is almost perfectly gaussian, and nearly power-law in frequency $P(\omega) \propto \omega^{-1.69}$ (e.g. Fig. \ref{fig1}).

\subsection{Present observations and prospects: Case study for SDSSJ1004+4112} \label{SDSS}
\begin{figure}[t]
\includegraphics[width=\linewidth]{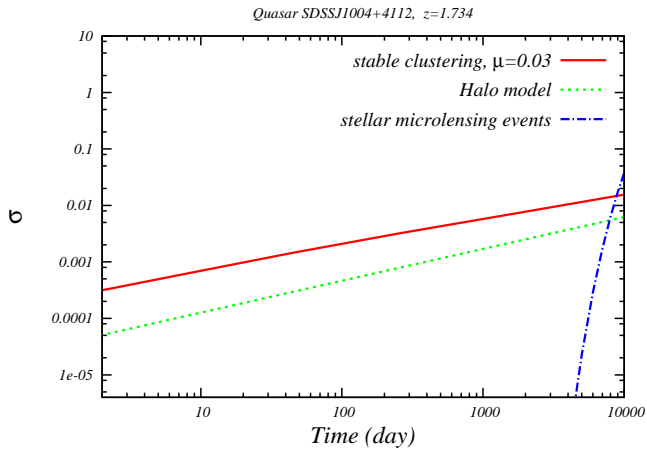}
\caption {The expected standard deviation of the magnification due to the transient weak lensing in the light curve of SDSSJ1004+4112 from one day up to $10^4$ days. The standard deviation is obtained for the stable
clustering (solid line) with $\mu=0.03$,   halo model fitting function
(dotted line) with the cut-off of $M_{min} = 10^{-6}M_{\odot}$. The expected signal from a typical  cosmological microlensing event is also shown for comparison (dot-dashed curve; see Appendix \ref{microlensing}).
}

 \label{figsigma}
\end{figure}

\begin{figure}[t]
\includegraphics[width=\linewidth]{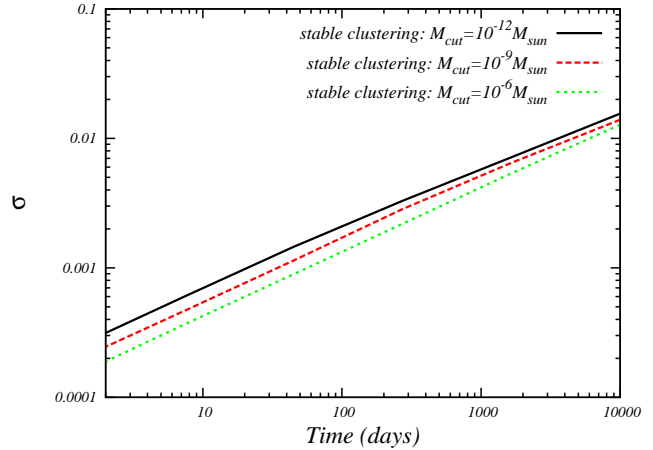}
\caption {Same as Fig. (\ref{figsigma}), but for different cutoff masses in the stable clustering model. Smaller cutoff provides 
larger variation in the light curve as the variance receives contribution from smaller haloes. This is a more significant effect for shorter time-scales, which are sensitive to smaller haloes.}
 \label{cutoff}
\end{figure}

In order to measure the light curve of a source with a given
photometric precision, we use the definition of the signal to noise
ratio in each data point due to deviation with respect to the base
line by $Q_i =  \delta A_i L_i/\sigma_L$, where $L_i$ is the
luminosity at $i$th point and $\sigma_L$ is the corresponding error
bar.  The variance of the signal to noise can be obtain by
$\sqrt{<Q^2>} =  \sqrt{<\delta A^2>} \bar{L}/\sigma_L$. In order to
have signal to noise larger than a given threshold, the error bar
should satisfy in the following constraint
\begin{equation}
\frac{\sigma_L}{\bar{L}}<\frac{\sqrt{<\delta A^2>}}{Q_{min}}.
\label{s2n}
\end{equation}
For a given source with the rate of $\beta$ photons per second
received by the telescope, the number of collected photons is $L =
\beta T_{exp}$. Considering photometric error due to the Poisson
fluctuations, the left hand side of equation (\ref{s2n}) can be written as
$(\beta T_{exp})^{-1/2}$, where the rate of photons received by a
telescope is given by $\beta = \frac{F}{<h\nu>}\pi (D/2)^2$. Then
the essential exposure time $T_{exp}$ to achieve a minimum signal to
noise of $Q_{min}$ is given by:
\begin{equation}
T_{exp}>\frac{Q_{min}^2}{<\delta A^2>}\frac{<h\nu>}{F_0}\frac{4}{\pi
D^2}10^{\frac{m-m_0}{2.5}}.
\label{const}
\end{equation}
Here we use the flux and magnitude of the Sun as reference with the
absolute magnitude of $m_0=4.75$ and corresponding photon flux of
$F_0/<h\nu> = 8.3 \times 10^8~ m^{-2}sec^{-1}$. For a small size
telescope with the diameter of $D=1.54$ m (as Danish Telescope at La
Silla), the minimum exposure time for a star with the magnitude of $m$, having an average signal to noise ratio larger than three, is
obtain by
\begin{equation}
T_{exp}(D = 1.54)>6.3\times 10^{-10}\times 10^{\frac{m-4.75}{2.5}}
\frac{Q_{min}^2}{<\delta A^2>} sec\label{const}.
\end{equation}
For the case of strong lensing images of SDSSJ1004+4112, we use the redshift of source at $z_s = 1.734$ \cite{oguri04}. Also, using the convergence and shear for the lensing clustering as  $\kappa = 0.392$ and $\gamma = 0.642$  \cite{Gordon}, we obtain the overall magnification for transient weak lensing using:
\begin{equation}
\label{mag2}
\frac{\delta A}{A_0}\simeq \frac{2 (\kappa^2+\gamma^2)^{1/2} \delta A_{int}}{|(1-\kappa)^2-\gamma^2|} \simeq 35\times\delta A_{int} .
\end{equation}
Here the magnification factor due to strong lensing is $A_0 = |(1-\kappa)^2-\gamma^2|^{-1} \sim 23.5$, and $ \delta A_{int}$ is the transient magnification, due to microhaloes, in the absence of the strong lens.  Figure (\ref{figsigma}) represents variance in the light curve of the Quasar from the transient weak lensing in terms of period of modes for two models of stable clustering, and Halo model fitting function. The variance in the fluctuation of the light curve depends on the cutoff on the mass of microhalos.  This cutoff 
in the mass is a function of micro-physics of dark matter particles that determines the size of free stream mass scale. Recent simulations provides the profile of the abundant microhalos \cite{mh} for the earth mass halos. In order to study the effect of cutoff mass of microhalos on transient weak lensing effect, we plot the variance in the light curve of the Quasar as a function time scale for three cases of cutoff masses  in Figure (\ref{cutoff}). Smaller cutoff results in larger variance in the variation of light curve by the transient weak lensing.

To achieve a $3\sigma$ signal to noise ratio, a variation in
magnification of $\sim 2\times 10^{-3}$ (which occur e.g. over a hundred days within the stable clustering model), Equation (\ref{const})
implies
\begin{equation}
T_{exp}(D=1.54)>1.14\times\times 10^{\frac{m-7.75}{2.5}}  sec .
 \label{const2}
\end{equation}

For the case of this Quasar, the magnitude of images are in the order of $m\simeq 18.5$, substituting in equation (\ref{const2}), we get the exposure time of $T_{exp}(D=1.54)>2\times 10^4$ sec. Increasing the
diameter of telescope reduces the exposure time as $T_{exp}\propto
1/D^2$. For a 10-meter class telescope, the exposure time reduces by
the factor of $0.023$ and we get the exposure time of
$T_{exp}(D=10)>500$ sec. Repeating this calculation for the halo model, which has a smaller standard deviation by an order of magnitude compared the stable clustering model, we would expect that  exposure time for detecting signal has to be two orders of magnitudes larger, i.e. $T_{exp}(\mbox{halo~model}) \sim 10^2 \times T_{exp}(\mbox{stable~clustering})$. 


From the technical point of view, we can achieve higher photometric accuracy with long term exposures even with the same size telescopes. One of the problems with the long exposure times is that CCD may saturate during the photometry. This problem can be solved
by telescope defocusing technique, which is widely used  in the
transit exoplanet observations \cite{Southworth}. Another solution
is that images can be taken with shorter exposures, then
stack to get a desired signal to noise ratio.

\section{Conclusion}
\label{conc} In this work, we propose the possibility of observation
of transient weak lensing of quasars, as a probe of the microhaloes
on the cosmological scales. These observations can rule out the
existence of dark microhaloes or constrain the properties of the
dark matter particles, as well as the models of non-linear structure formation. Assuming different models for non-linear structures on small
scales, we formulate the weak lensing of the microhaloes by the
geodesic and angular distance methods. We note that, due to weak
gravitational potential of these structures they don't host
baryonic/stellar matter in their centers.

The transient weak lensing signal is caused by random contributions of millions of microhaloes, and thus yields an almost perfect gaussian photometric noise.
We obtained the temporal power spectrum of this photometric noise in the
light curve, finding a near power-law red spectrum of $P(\omega) \propto \omega^{-1.69}$. For the
quasar strong lensing systems, the variation in the light curve due to transient weak lensing can be of the order of $10^{-4}-10^{-3}$, depending on the halo models.

In contrast,  the intrinsic  variations of quasar light can be as much as an order of magnitude. However, we have argued that,
in principle, we can separate the intrinsic variation of quasars from the transient weak lensing by
the time delay method for multiple images in a strong lensing systems. Moreover, we showed that the primary contamination of this signal by stellar microlensing can be effectively cleaned, as it has a significantly different amplitude, statistics, and frequency dependence.



Finally, for the observation of transient weak lensing we proposed monitoring images of quasars in a strong lensing system, preferably lensed with a cluster of galaxies. Amongst bright lensed Quasars in the SDSS Data Release 7,  SDSSJ1004+4112 is lensed by cluster of galaxy with images formed at large angle separation. The advantage of using strong lensing system is that (i) we can remove the intrinsic variations of a quasar by time delay method and (ii) the transient weak lensing signals can be enhanced by the magnification of the source due to the strong lensing.  For the  case of stable clustering, we calculated an exposure time of $\sim 500$ sec with a 10-meter class telescope to achieve $3\sigma$  signal to noise ratio in detection of time variation by the transient weak lensing. A long term monitoring of this system is therefore proposed as a means to study the nature of dark matter microhaloes on the cosmological scales. The result of observations, either will constrain the existence of microhaloes, or rule them out as the building blocks of dark matter structures in favour of modified gravity alternatives, or more conservatively, a breakdown in CDM hierarchy on small scales.

%

%

\begin{acknowledgements}
We would like to especially thank Avery Broderick for reminding us about the power of multiple-image strong lensing for differentiating our signal from intrinsic quasar variability. We  would also like to thank Adrienne Erickcek, Neal Dalal, and James Taylor for invaluable discussions and/or feedback on the manuscript. This research was supported by Perimeter Institute for Theoretical
Physics and the John Templeton Foundation. Research at Perimeter
Institute is supported by the Government of Canada through Industry
Canada and by the Province of Ontario through the Ministry of
Economic Development \& Innovation.

\end{acknowledgements}

\appendix
\section{Transient Weak Lensing: Angular Diameter-Distance method}
\label{angd}

For a bundle of light rays emitting from a source, the relative
change on the surface of light bundle in a generic metric is given
by
\begin{equation}\label{dA}
\frac{dA}{Ad\lambda}=2\theta=k^{\mu}{}_{;\mu},
\end{equation}
where $k^{\mu}$ is the tangent vector to the null geodesics and
$\theta$ represents the isotropic contraction and expansion of the
light beam. The equation governs the dynamics of $\theta$ and shear
is given by \cite{Schneider1992}:
\begin{equation}\label{thetadot}
{\theta}^{\prime}+\theta
^2+|\sigma|^2=-\frac{1}{2}R_{\mu\nu}k^{\mu}k^{\nu},
\end{equation}
where $\prime$ stands for derivative with respect to the affine
parameter and $|\sigma|$ is the magnitude of shear tensor define as
below:
\begin{equation}
|\sigma|^2=\frac{1}{2}k_{\mu ;\nu}k^{\mu ;
\nu}-\frac{1}{4}(k^{\alpha}_{~;\alpha})^2. \label{shear}
\end{equation}
Combining equations (\ref{dA}) and (\ref{thetadot}) results in the
focusing equation
\begin{equation}\label{focus}
{\sqrt{A}}^{\prime\prime}=-(|\sigma|^2+\frac{1}{2}R_{\mu\nu}k^{\mu}k^{\nu})\sqrt{A}.
\end{equation}
This equation can be used for studying the influence of matter on
the light bundle while it propagates from the source to the
observer. The effect of matter distribution relates to the Ricci
tensor and the magnitude of shear tensor. The components of Ricci
tensor from the metric in equation (\ref{pert-met}) is given by
\cite{Dodelson2003}:
\begin{equation}
R_{00}=-3\frac{\ddot{a}}{a}-a^{-2}\nabla
^2\Phi-\ddot{\Phi}+6H\dot{\Phi},
\end{equation}
and
\begin{equation}
R_{ij}=\delta_{ij}\left[(2a^2H^2+a\ddot{a})(1+4\Phi)+7a^2H(\dot{\Phi})+a^2\ddot{\Phi}-\nabla
^2\Phi\right],
\end{equation}
where "~$\dot{}$~" is derivative with respect to coordinate time,
 $t$. For solving the focusing equation we replace the $k^\mu =
{dx^\mu}/{d\lambda}$ with the observable parameters. Here $\lambda$
is defined similar to Eq.(\ref{affine}). We set $\lambda$ to zero at
the position of the observer and the finial value at the location of
the source, then $adt=a^2d\chi = -d\lambda$ and  $k^\mu$ in FRW
metric can be given by:
\begin{eqnarray}
k^{\mu}&=&(-a^{-1},-a^{-2},0,0),\\ \nonumber k_{\mu}&=&(
a^{-1},-1,0,0).
\end{eqnarray}
On the other hand the Ricci term in Eq.(\ref{focus}) is obtained as
\begin{equation}\label{riccitensor}
\frac{1}{2}R_{\mu\nu}k^{\mu}k^{\nu}=\frac{4\pi
G\rho}{a^2}+\frac{16\pi G\rho}{a^2}\Phi-\frac{1}{a^4}\nabla
^2\Phi-\frac{\ddot{\Phi}}{a^2}-\frac{H\dot{\Phi}}{a^2},
\end{equation}
For the shear term we need to calculate the following elements:
\begin{equation}
k_{\mu ;\nu}k^{\mu ;
\nu}=k_{0;0}k^{0;0}+k_{i;0}k^{i;0}+k_{0;i}k^{0;i}+k_{i;j}k^{i;j}
\end{equation}
where $i,j$ represent spatial components and
\begin{eqnarray}\label{sigmacomp}
k_{0;0}k^{0;0}&=&(\frac{H}{a})^2+\frac{2H}{a^2}(-\dot{\Phi}+\frac{\Phi
_{,x_{||}}}{a}), \\ \nonumber
k_{i;0}k^{i;0}&=&k_{0;i}k^{0;i}=-(\frac{H}{a})^2-2Ha^{-2}(\dot{\phi}+a^{-1}\Phi_{,x_{||}}),
\\ \nonumber
k_{i;j}k^{i;j}&=&3(\frac{H}{a})^2+24(\frac{H}{a})^2\Phi+6\frac{H\dot{\Phi}}{a^2}+\frac{2}{a^3}H\Phi_{,x_{||}},
\end{eqnarray}
Substituting in equation (\ref{shear}), the shear term obtain as
follows:
\begin{equation}\label{shea}
|\sigma|^2=12(\frac{H}{a})^2\Phi-\frac{2H\dot{\Phi}}{a^2}-\frac{2H}{a^3}\Phi_{,x_{||}}.
\end{equation}
Finally, substituting Equations (\ref{riccitensor}) and (\ref{shea})
in focusing equation and changing the affine derivatives to the
coordinate time derivatives, we get:
\begin{equation}
\ddot{D}-H\dot{D}+4\pi\rho G D+{\cal{F}}D=0,
\end{equation}
where we rename $D\equiv\sqrt{A}$, representing the angular diameter
distance to any point at the light bundle and ${\cal{F}}$ is given
by:
\begin{equation}
{\cal{F}}=15H^2\Phi-3H\dot{\Phi}-a^{-2}\nabla
^2\Phi-\ddot{\Phi}-2Ha^{-1}\Phi_{,x_{||}}.
\end{equation}
Note that some of these terms with the factor of $H^{-1}$ represent
variation of the perturbation during the Hubble time scale, some
terms with the partial derivative with respect to the time
$\partial_t$ represent intrinsic time variation of the perturbations
and some terms with $\partial_x$ represents the spatial variation of
the field. In order to get a differential equation for perturbation
of $D$, we write the angular diameter distance as sum of $D_{0}$ due
to the unperturbed FRW and the first order perturbation term $D_1$:
\begin{equation} D=D_{0}+D_1.
\end{equation}
The background evolution and the first order perturbation can be
divided into two equations as:
\begin{eqnarray}\label{bg}
\ddot{D}_{0}-H\dot{D}_{0}+4\pi G\rho D_{0}=0, \\
\label{deltaD} \ddot{D_1}-H\dot{D_1}+4\pi G \rho D_1+
{\cal{F}}D_{0}=0,
\end{eqnarray}
where the solution of Eq. (\ref{bg}) in terms of the comoving
distance is $D_{0} = a \chi$. To solve the Eq.(\ref{deltaD}), we
assume that the time variation due to the motion and spatial
gradient of substructures is smaller than the Hubble time scale
\begin{equation}
{\Delta x}/c, t_{DM}\ll t_{Hubble},
\end{equation}
hence $H\dot{\Phi}\ll \ddot{\Phi}$, $H^2\Phi\ll \ddot{\Phi}$,
$H/a\Phi_{,x||}\ll\ddot{\Phi}$ and Eq.(\ref{deltaD}) reduces to:
\begin{equation}\label{deltaD1}
\ddot{D_1}-\left[\ddot{\Phi}+a^{-2}\nabla ^2\Phi(x,t)\right]D_{0}=0.
\end{equation}
We use Fourier transform for $\Phi$ and consider a transverse
velocity for the perturbation as $x_\bot \rightarrow x_\bot +
\frac{v_\bot t}{a}$. Hence time derivative of $\Phi$ is given by
\begin{equation}
\ddot{\Phi}(\chi,t)=-\int
\frac{d^3k}{(2\pi)^3}(\frac{k_{\bot}v}{a})^2\Phi(k)e^{-ik_{||}\chi_{||}}e^{-ik_{\bot}x_{\bot}(t)},
\end{equation}
and for the gradient of perturbation we have
\begin{equation}
\frac{1}{a^2}\nabla ^2\Phi(\chi,t)=- \int
\frac{d^3k}{(2\pi)^3}\frac{k_{||}^2+k_{\bot}^2}{a^2}
\Phi(k)e^{-ik_{||}\chi_{||}}e^{-ik_{\bot}x_{\bot}(t)}.
\end{equation}
Assuming that $v\ll1$, we can ignore the time derivative term of
$\Phi$, then Eq.(\ref{deltaD1}) reduces to
\begin{equation}\label{deltaD12}
\ddot{D_1}(t)=-D_{0}(t) \int \frac{d^3k}{(2\pi)^3}
\frac{k_{\bot}^2+k_{||}^2}{a^2}\Phi(k)e^{-ik_{||}\chi_{||}}e^{-ik_{\bot}x_{\bot}(t)}.
\end{equation}
For a perturbed FRW metric in flat universe, the angular diameter
distance can be written as $D=a(\chi+\epsilon(\chi))$, where
$\epsilon(\chi)$ is the first order perturbation in terms of
comoving distance and $D_1 = a \epsilon(\chi)$. Hence the
perturbation in the luminosity distance along the propagation of the
light can be written as:
\begin{equation}\label{epsilon}
\ddot{D_1}=\frac{1}{a}\frac{d^{2}\epsilon(\chi)}{d\chi ^2}.
\end{equation}
Consequently equation(\ref{epsilon}) can be written as:
\begin{widetext}
\begin{equation}
\epsilon(\chi)=-\int \frac{d^3k}{(2\pi)^3}
({k_{\bot}}^2+k_{||}^2)\Phi(k)e^{-ik_{\bot}\chi_{\bot}}\int_{0}^{\chi_{s}}d\chi^{\prime}\int_{0}^{\chi
^{\prime}}e^{-ik_{||}\chi^{\prime\prime}} \chi^{\prime\prime}d\chi
^{\prime\prime}.
\end{equation}
We can simplify this integral as
\begin{equation}
\epsilon(\chi)=-\chi_{s}\int\frac{d^3k}{(2\pi)^3}
({k_{\bot}}^2+k_{||}^2)\Phi(k)e^{-ik_{\bot}\chi_{\bot}}\int_{0}^{\chi_{s}}
e^{-ik_{||}\chi ^{\prime}}\chi ^{\prime}(1-\frac{\chi
^{\prime}}{\chi_s})d\chi ^{\prime}.
\end{equation}
Now we can calculate the relative perturbed angular distance to the
background angular distance as follows
\begin{equation}
\frac{\delta D}{D_{0}}=-\int_{0}^{\chi_{s}}\int
\frac{d^3k}{(2\pi)^3}(k_{\bot}^2
+k_{||}^2)\Phi(k)e^{-ik_{\bot}\chi_{\bot}} e^{-ik_{||}\chi
^{\prime}}\chi ^{\prime}(1-\frac{\chi ^{\prime}}{\chi_s})d\chi
^{\prime}.
\end{equation}
\end{widetext}
In this integral, the longitudinal modes with wave length smaller
than $\chi_s$ can cancel the effect of each other and only
$k_{||}<\chi_s^{-1}$ modes have non-zero contribution in the
integration. On the other hand in order to have short transient
events the width of modes must be much smaller than the longitudinal
size, $k_{||}\ll k_{\bot}$. Since the flux of source at the position
of observer is proportional to the inverse of square of angular
distance, $F\propto D^{-2}$, the magnification in terms of angular
distance can be written as:
\begin{equation}
A=\frac{F}{F_{0}}=(1+\frac{D_1}{D_{0}})^{-2}.
\end{equation}
So perturbation in the magnification obtain as
\begin{equation}
\delta A=2\int_{0}^{\chi_{s}}\int \frac{d^3k}{(2\pi)^3}
k_{\bot}^2\Phi(k)e^{-ik_{\bot}\chi_{\bot}}e^{-ik_{||}\chi
^{\prime}}\left(1-\frac{\chi ^{\prime}}{\chi}\right)\chi ^{\prime}d\chi
^{\prime}, \label{deA}
\end{equation}

This expression is identical to what is obtained from the geodesics
method in equation (\ref{permag}). We note that $\delta A$ is a time
dependent function as the transverse position of the structures
changes with respect to our line of sight.

\section{Temporal power spectrum of microlensing}\label{microlensing}

\begin{figure}[t]
\includegraphics[width=\linewidth]{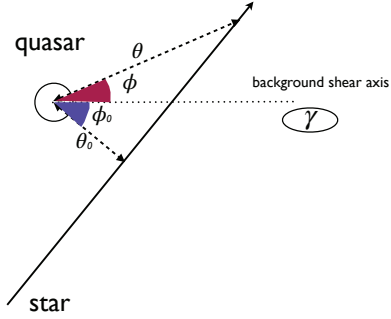}
\caption {Schematics of stellar microlensing in background shear field. 
}

 \label{microlensing_fig}
\end{figure}

In this appendix, we compute the expected contribution to the magnification temporal power spectrum due to stellar microlensing events. 
We first note that the magnification $A$ is given by:
\beq
A^{-1} =  A^{-1}_0+2 \gamma (\theta_E/\theta)^2 \cos(2\phi) - (\theta_E/\theta)^4,\label{a_inv}
\eeq
where $A_0$ and $\gamma$ are magnification and shear of the smooth background. $\phi$ is the angle between the principal axis of the background shear field, and the line that connects star to the quasar in the sky. For a star moving on a straight line, with impact parameter $\theta_0$ (see Fig. \ref{microlensing_fig}) and proper motion $\dot{\theta}$, the Fourier transform of  Eq. (\ref{a_inv}) is:
\begin{widetext}
\beq
A^{-2}_0\delta A_\omega \simeq \frac{\pi\theta^2_E}{\dot{\theta}\theta_0}e^{-i\omega t_0-|\omega \theta_0/\dot{\theta}|}\left[\frac{\theta^2_E}{2\theta^2_0}(1+|\omega \theta_0/\dot{\theta}|)+2\gamma |\omega \theta_0/\dot{\theta}| e^{2i\phi_0 \cdot{\rm sgn}(\omega)}\right],
\eeq
where $t_0$ and $\phi_0$ are time and $\phi$ when $\theta=\theta_0$. To find the total temporal power spectrum of magnification, we have to add the $\delta A_\omega$'s from all the stars. However, assuming a random distribution of stars, the contributions only add in quadratures:
\beq
P_{\rm stars}(\omega)= A^4_0T_{obs}^{-1} \sum_{\rm stars}  \frac{\pi^2\theta^4_E}{\dot{\theta}^2\theta_0^2}e^{-2|\omega \theta_0/\dot{\theta}|}
\left|\frac{\theta^2_E}{2\theta^2_0}(1+|\omega \theta_0/\dot{\theta}|)+2\gamma |\omega \theta_0/\dot{\theta}| e^{2i\phi_0}\right|^2,\eeq
or
\bea
\sigma_{\rm stars}^2 &\equiv& \int_{ T^{-1}_{obs}}^\infty P_{\rm stars}(\omega) \frac{d\omega}{2\pi} \nonumber\\ &=& \sum_{\rm stars} \frac{ \pi A^4_0\theta^{4}_E}{32\theta_0^7\dot{\theta}^3T^3_{obs}} \exp\left(-\frac{2\theta_0}{\dot{\theta} T_{obs}}\right) \left[2 (\theta_0^2 \theta^{4}_E+ 16 \gamma^2 \theta_0^6) + 
  2 \theta_0 (3\theta^{4}_E + 16 \gamma^2 \theta_0^4) \dot{\theta} T_{obs} + (5 \theta^{4}_E+ 
     16 \gamma^2 \theta_0^4) \dot{\theta}^2 T_{obs}^2\right]\nonumber\\
\eea
\end{widetext}
where we have suppressed  the dependence of the parameters $\theta_E, \theta_0,\phi_0$ and $\dot{\theta}$ on the star, and further used $\langle e^{i\phi_0} \rangle =0$ in the last equality. 
	
Therefore, we see that typical events with $t(\theta) \sim \frac{\theta_0}{\dot{\theta}} \sim 10^2$ yr's (Eq. \ref{t_theta}), are suppressed by exponential of $t(\theta)/T_{obs} \sim 10^2$ over relevant time-scales of $T_{obs} \sim$ yr, and thus are completely negligible.

\end{document}